\documentclass[12pt]{article}
\usepackage{amsmath}
\usepackage{graphicx}
\usepackage{amssymb}

\usepackage{hyperref}
\renewcommand{\title}[1]{%
    \bigskip
    \begin{center}%
    \Large\bf #1%
    \end{center}%
    \vskip .2in}

\renewcommand{\author}[1]{%
    {\begin{center}
    #1
    \end{center}}}
\newcommand{\address}[1]{\vspace{-1.7em}\vspace{0pt}
    {\begin{center}
    \it #1
    \end{center}}}

\begin{document}

%\twocolumn[
 % \begin{@twocolumnfalse}

\title{Galilean gauge theory from Poincare gauge theory }

\author
{
Rabin Banerjee  $\,^{\rm a,b}$,
%Arpita Mitra    $\,^{\rm a, e}$,
Pradip Mukherjee $\,^{\rm c,d}$}\vspace{0.5em}
 %\footnote{Also, Visiting Associate, S. N. Bose National Centre for Basic Sciences, JD Block, Sector III, Salt Lake City, Kolkata -700 098, India  } 
%,
\address{$^{\rm a}$S. N. Bose National Centre 
for Basic Sciences, JD Block, Sector III, Salt Lake City, Kolkata -700 106, India }\vspace{0.5em}

\address{$^{\rm c}$Department of Physics, Barasat Government College,\\10, KNC Road, Barasat, Kolkata - 700124, India.

 }\vspace{0.5em}

\address{$^{\rm b}$\tt rabin@bose.res.in}\vspace{0.5em}
\address{$^{\rm d}$\tt mukhpradip@gmail.com}
%\address{$^{\rm e}$\tt arpita12t@bose.res.in}

\begin{abstract}
We provide an exact mapping between  the Galilian gauge theory, recently advocated by us \cite{BMM1, BMM2, BM}, and  the Poincare gauge theory. Applying this correspondence  we provide a vielbein approach to the geometric formulation of Newton's gravity where no ansatze or additional conditions are required.
\end{abstract}
\section{Introduction} In the broad physics community geometric theory of gravity usually means general relativity (GR). But geometrical formulation is nothing special of relativistic gravity. The fact is that, almost simultaneously with Einstein, Cartan \cite{Cartan-1923, Cartan-1924} recast Newtonian gravity on Newton Cartan (NC) spacetime.   
It was subsequently developed  \cite{Havas} -\cite{MALA} by many stalwarts. The aim of such studies was to formulate Newtonian gravity covariantly. 

A different type of problems emerged in the recent past where it was required to couple a field theory having Galilean symmetry with background gravity \cite{SW}. Besides other issues, 
there were problems of anomalous transformations of the metric and the flat limit could not be consistently imposed\cite{BGM}. In this perspective we introduced the Galilean gauge theory\cite{BMM1}, \cite{BMM2}, which provided a systematic algorithm for coupling a field theory with nonrelativistic gravity. The method was inspired by Utiyama's idea of gauging the Poincare symmetry of a field theory \cite{blago} which later developed as the well known Poincare gauge theory (PGT). We have discussed many contemporary problems as an application of GGT  \cite{BGM},\cite{BM3},\cite{RBPMt} ,\cite{BMM3} ,
always  getting physically consistent results. Just as in PGT gauging  Poincare symmetry leads to Riemann Cartan spacetime,GGT gauges the (extended) Galilean symmetries to lead to Newton Cartan spacetime. Since in flat space time Galilean invariant theories may be obtained from Poincare invariant theories by an appropriate reduction \cite{last}, one expects a similar situation in curved background also. Thus there should be a  connection between PGT
and GGT. Since there are many intricasies when gravity is invoolved \cite{BGM} such connections are always worthy to be studied.

 We show in this paper that there exists an exact mapping between PGT and GGT. Apart from  theoretical satisfaction, such a mapping has immense practical value of finding the non relativistic limit of a theory of gravity. We
demonstrate this by deriving Newtonian gravity from general relativity without any other approximation or ansatze. This is a new vielbein approach to the old problem of formulating the Newtonian gravity covariantly using the Newton-Cartan metrics.

  After this introduction we review in section 2 the rudiments of Newton-Cartan gravity. In section 3, the basic structure of GGT for a generic field theory is discussed. The transformation relations applicable to a field which is in  an arbitrary representation of rotation group in three dimensions are obtained. This is an extension of previous results \cite{BMM1}. The emergence of the Newton Cartan
  spacetime is demonstrated. In the next section we discuss how GGT may be regarded as the nonrelativistic limit of PGT. 
  The transformation equations derived in section 3 are obtained from the corresponding transformations of PGT by
% imposing appropriate conditions on the 
working out the appropriate dictionary involving the field components and transformation parameters. The conditions on the field components are the same as those we have assumed on physical considerations in constructing GGT. This is remarkable and points out to an inner harmony of GGT. In section 5 the correspondences obtained in section 4 are used to derive the complete set of equations of Newtonian gravity from Einstein's gravity. We first use the expansions of the metric and metric compatible symmetric connection in terms of the vielbeins and spin connections. Then the identity of the vielbeins and spin connections with the Poincare gauge fields are utilised. Since already we have proved the correspondence of PGT and GGT, directly we can write corresponding relations in NC spacetime. The Einsteinian dynamics is then reduced to NC manifold which is
%under appropriate conditions. This is 
just the Newtonian gravity, as we will see. We conclude in section 6.

 \section{Newton Cartan gravity}  It was Cartan who first provided a geometric
formulation of Newtonian gravity. Apart from yielding a novel geometry named after
Newton-Cartan (NC), this analysis gave an essentially new and deep insight into
Einstein's original formulation of GTR.

Newton's theory is encapsulated in the trajectory of neutral test particles
\begin{equation}
\frac{d^2 x^i}{d t^2} + \frac{\partial \phi}{\partial x^i} = 0
\label{ntp.eom}
\end{equation}
where $x^i (i = 1,2,3)$ are the spatial coordinates and the source equation for the
Newtonian potential $\phi$ is given by
\begin{equation}
\nabla^2 \phi = 4 \pi \rho G
\label{poi.eqn}
\end{equation}
with $\rho$ being the mass density.

In standard Newtonian interpretation the above equations describe a curved
trajectory in flat three dimensional space. Cartan generalized this viewpoint by
interpreting the trajectories as geodesics in four dimensional curved spacetime,
\begin{equation}
\frac{d^2 x^{\mu}}{d t^2} + \Gamma^{\mu}_{\nu \rho} \frac{d x^{\nu}}{dt} \frac{d
x^{\rho}}{dt} = 0
\label{geo.eqn}
\end{equation}
This is possible if one takes $x^{\mu} = (x^0 = t, x^i)$ and chooses the ansatz 
\begin{equation}
\Gamma^i_{0 0} = \frac{\partial \phi}{\partial x^i} \quad \,, \quad \text{all other}
\, \Gamma^{\mu}_{\nu \rho} \, \text{vanish} 
\label{chr.sym}
\end{equation}
Inserting this in the standard expression for the Riemann tensor
\begin{equation}
R^{\alpha}_{\phantom{\alpha} \beta \gamma \delta} = \partial_{\gamma}
\Gamma^{\alpha}_{\beta \delta} - \partial_{\delta} \Gamma^{\alpha}_{\beta \gamma} +
\Gamma^{\alpha}_{\mu \gamma} \Gamma^{\mu}_{\beta \delta} - \Gamma^{\alpha}_{\mu
\delta} \Gamma^{\mu}_{\beta \gamma}
\label{rie.ten}
\end{equation}
one finds
\begin{equation}
R^i_{0j0} = \partial_i \partial_j \phi \quad \,, \quad \text{all other} \,
R^{\alpha}_{\beta \gamma \delta} \, \text{vanish}
\label{rie.com}
\end{equation}
Then Poisson's equation (\ref{poi.eqn}) is expressed as,
\begin{equation}
R_{00} = 4 \pi \rho G
\label{ric.com}
\end{equation}
Equations (\ref{geo.eqn}) to (\ref{ric.com}) summarize the geometric formulation of
Newton's gravity.

Let us now introduce the elements of NC geometry. The special role of time in the NR
case is manifested by the lack of a single nondegenerate metric. Rather there are
two degenerate metrics given by a temporal one-form $\tau_{\mu}$ and a degenerate
second rank metric $h^{\mu \nu}$ of rank 3 so that,
\begin{equation}
h^{\mu\nu}\tau_{\nu} = 0 
\label{ort.rel}
\end{equation}
There is a vector $v^{\mu}$ corresponding to $\tau_{\mu}$ such that,
\begin{equation}
v^{\mu}\tau_{\mu} = 1
\label{tau.nor}
\end{equation}
Also, there is a second rank covariant tensor $h_{\mu\nu}$ which is introduced to
project any vector to a space like one form satisfying,
\begin{equation}
h_{\mu\nu}v^{\nu} = 0
\label{ort.rl2}
\end{equation}
while the projection operator is
\begin{equation}
P^{\rho}_{\mu} = h_{\mu \nu}h^{\nu \rho} = \delta^{\rho}_{\mu} - \tau_{\mu} v^{\rho}
\label{pro.opt}
\end{equation}

Finally, $h^{\mu\nu}$ and $\tau_\mu$ are assumed to be covariantly conserved with
respect to the affine connection $\Gamma_{\mu\nu}^\alpha$,
\begin{equation}
\nabla_{\alpha}h^{\mu\nu} = \nabla_{\alpha}\tau_{\nu} = 0
\label{ort.rl}
\end{equation}

Imposing certain conditions and choosing an ansatz \cite{Daut, TrautA, EHL}, relations (\ref{chr.sym}) 
and (\ref{ric.com}) are reproduced, thereby yielding Newton's gravity.

 \section{Galilean gauge theory}
We provide here a short account of deduction of the structure and geometrical interpretation of GGT, when the matter field is in any arbitray representation of the rotation group in three space. The existing results \cite{BMM1,BM} were derived for scalar field. The procedure for the more general case mimics the old one, the only change is in the expansion of the gauge field in terms of the spin connections.
The new features appearing due to the generalisation will be indicated at the appropriate places.

% Apart from this generalisation the purpose is to acquaint the reader with our notations and to compile the necessary transformation rules of the basic fields which will be required subsequently..
\subsection{Formulation of Galilean gauge theory}
  A non-relativistic theory is assumed to be given by the action,
\begin{equation}
S=\int dx^0d^3x\cal{L}(\phi, {\bf \nabla}\phi)\label{gc}
\end{equation}
where $\phi$ is an arbitrary field  in 3-dim Eucledian space 
The most general infinitesmal coordinate transformation consistent with Galileo Newton conceptions about space and time are:
\begin{equation}
x^\mu \to x^\mu + \zeta^\mu ; \zeta^0 =  -\epsilon, \zeta^k = \eta^k - v^k t; \eta^k= \omega^k{}_l + \epsilon^k\label{gtrans} 
\end{equation}
where $\omega^k{}_l$ is the 3-dim rotation parameters, $-\epsilon,\epsilon^k $ are
parameters for time and space translations and $v^k$ are  the galilean boosts. %parameter. Specialising for infinitesmal  transformations 

The transformations of the fields are given by,
%Defining the form variation as $\delta_0\phi =\phi'(x) -  \phi(x)$, we get
\begin{equation}
\delta_0\phi=-\zeta^{\nu}\partial_{\nu}\phi+\frac{1}{2}\omega^{\alpha\beta}\sigma_{\alpha\beta}\phi-imv^ax_a\phi\label{Delphi}
\end{equation}
where $\delta_0\phi =\phi'(x)-  \phi(x)$ denotes the form variation .

Differentiating (\ref{Delphi}) and utilising the commutative property of differentiation and form variation ($\delta_0\partial_k\phi = \partial_k\delta_0\phi$), we get the form variations of the time and the space variables as,
\begin{eqnarray}
&\delta_0 \partial_k\phi=-\xi^{\mu}\partial_{\mu}(\partial_{k}\phi)-imv^ix_i\partial_{k}\phi-\lambda^{m}{}_{k}\partial_{m}\phi-imv_k\phi\nonumber\\
&\delta_0 \partial_0\phi= -\xi^\mu\partial_{\mu}(\partial_{0}\phi)-imv^i x_i\partial_{0}\phi+v^{i}\partial_{i}\phi
\label{delkphin}
\end{eqnarray}
It is notable that the transformations (\ref{Delphi}) and (\ref{delkphin}) ensure uhe invariance of the generic theory (\ref{gc})
under the global galilean transformation (\ref{gtrans}).

To localise the symmetry we require to set up local coordinates with respect to which the local galilean transformations are defined. In flat space, the relation between the local and spacetime coordinates is \footnote{Latin indices from the beginning $\left(a, b,....\right)$
denote the local basis while the coordinate basis is defined from the middle $\left(k, l,....\right)$}:
\begin{equation}
{\bf{e}}_a = \delta_a^k{\bf{e}}^k
\label{basis}
\end{equation}
This apparently trivial status of the local coordinates is dramatically altered when we invesigate the geometric connection.

When the galilean transformation parameters are functions of space time the partial derivatives $\partial_k\phi$ and $\partial_t\phi$ no longer transform as (\ref{delkphin}). Following the gauge procedure one needs to introduce covariant derivatives which will transform  as (\ref{delkphin}). Our experience with PGT indicates that this covariant derivative has to be constructed in two steps. The first step in the process of localisation is to convert the ordinary derivatives into covariant derivatives with respect to the global coordinates. Let us introduce the gauge fields $B_0$ and $B_k$ such that,
% to define the covariant derivatives with respect to the global coordinates
\begin{eqnarray}
D_k\phi=\partial_k\phi+iB_k\phi\nonumber\\
D_0\phi=\partial_0\phi+iB_0\phi \label{firstcov1}
\end{eqnarray}
The gauge fields $B_0$ and $B_k$ correspond to gauging the rotations and galilean boosts.They have the structures,
\begin{equation}
B_{\mu}=\frac{1}{2}B_\mu{}^{ab}\sigma_{ab}+\frac{1}{2}B_\mu{}^{a}(mx_{a})\label{gaugefields1}
\end{equation}
where $\sigma_{ab}$ and $\sigma _{a}$ are respectively the generators of rotations and Galileo boosts. 
%Global covariant derivatives are not the final covariant detivatives.

In the second step we obtain the true  covariant derivatives, defined with respect to he local coordinates, as, 
\begin{eqnarray}
\nabla_a\phi&=&{\Sigma_a}^{k}D_k\phi\nonumber\\
\nabla_{0}\phi&=&({\Sigma_0}^{0}D_0 \phi+{\Sigma_0}^{k} D_k\phi)\label{A}
\end{eqnarray}
where ${\Sigma_a}^{k}$ and ${\Sigma_0}^{\mu}$ are new fields introduced due to
localisation. Observe that $\Sigma_a{}^0=0$. This is connected with the specific properties of the galilean transformations \cite{BMM1}.

 Following the methodology of GGT we find that the invariance of the theory is retained after localization if the newly introduced  basic fields transform as 

\begin{eqnarray}
 \delta_0\Sigma_0{}^0 &=& -\zeta^{\nu}\partial_{\nu} \Sigma_0{}^0+ \Sigma_0{}^{\nu}\partial_{\nu}\zeta^0\notag\\ \delta_0\Sigma_0{}^{k} &=& -\zeta^{\nu}\partial_{\nu} \Sigma_0{}^k+\Sigma_0{}^{\nu}\partial_{\nu}\zeta^{k}+u^a
\Sigma_a{}^k\notag\\ \delta_0\Sigma_a{}^k &=& -\zeta^{\nu}\partial_{\nu}\Sigma_a{}^k+\Sigma_a{}^{\nu}
\partial_{\nu}\zeta^k- \omega_a{}^b\Sigma_b{}^k \notag\\\delta_0 B_\mu &=& -\zeta^{\nu}\partial_{\nu}B_\mu
-\partial_\mu\zeta^{\nu} B_{\nu}-\frac{1}{2}\partial_{\mu}\omega^{ab}\sigma_{ab}
+\partial_{\mu}u^a{x_a}
          \label{crap1} 
 \end{eqnarray}
%\lable{crap1} 

We can work out the form variation $\delta_0 B_{\mu}$ from (\ref{gaugefields1}) where $\delta_0B_\mu{}^{ab}$ and $\delta_0B_\mu{}^{a}$ will appear as unknowns. Alternatively, $\delta_0 B_{\mu}$ is given by (\ref{crap1}). Combining the two expressions of the same quantity,  we get,
\begin{align}
\delta_0 B_{\mu}^{ab} &= - \omega^a_c B_{\mu}^{cb} - \omega^b_c B_{\mu}^{ac} -\partial_{\mu} \omega^{ab} - \partial_{\mu} \zeta^{\nu}B_{\nu}^{ab} - \zeta^{\nu}\partial_{\nu}B_{\mu}^{ab} \notag\\
\delta_0 B_{\mu}^{a} &= - \omega^a_b B_{\mu}^{b} + \partial_{\mu} u^a - \partial_{\mu} \zeta^{\nu}B_{\nu}^{a} - \zeta^{\nu}\partial_{\nu}B_{\mu}^{a} \label{crap}
\end{align}

 We thus provide an algorithm for localising the Galilean symmetry of a nonrelativistic model which  is now elaborated. Introduce local coordinates at each point of 3-d space  as replica of the spacetime coordinate system. The local basis is connected to the coordinate basis by (\ref{basis}). If the original theory is given by the action (\ref{gc}) which is invariant under the global Galilean transformation
 defined in (\ref{gtrans}), then
\begin{equation}
S = \int dx^{\bar{0}} d^3x \frac{\det{\Sigma}^{-1}}{\Sigma_0{}^0}{\cal{L}}\left(\phi, \nabla_{\bar{0}}\phi, \nabla_a\phi\right)
\label{localaction}
\end{equation}
is invariant under the corresponding local Galilean transformations. Note that a correction factor to the measure is included as the jacobian
$\partial_\mu\zeta^\mu$ is not equal to zero now. The above result has been explicitly derived in (\cite{BMM1, BMM3}).

\subsection{Geometric connection}

The modified theory (\ref{localaction}) has a geometrical interpretation, The fields ${\Sigma_a}^{k}$ and ${\Sigma_0}^{\mu}$ may be reinterpreted as vielbeins in a general manifold charted by the  coordinates $x^{\mu}$ connecting the spacetime coordinates with the local coordinates that represent the local Galilean symmetry. From this viewpoint the connection between the spacetime basis and the local basis will be nontrivial (in contrast to (\ref{basis}))
\begin{equation}
{\bf {e}}_a = {\Sigma_a}^k{\bf{e}}_k
\label{nbasis}
\end{equation}
It has been proved that the 4-dim spacetime obtained in this way above is the Newton-Cartan
manifold. This is done by showing that the metric formulation of our theory contains the same structures and satisfy the same structural relations as in NC space - time \cite{BMM2}. Indeed the various elements of the NC geometry introduced in section 2 are given in terms of the GGT variables by,
   
 \begin{equation}
h^{\mu\nu}={\Sigma_a}^{\mu}{\Sigma_a}^{\nu}; \hspace{.2cm}\tau_{\mu}={\Lambda_\mu}^{0}
\label{spm}
\end{equation}
and,
\begin{equation}
h_{\nu\rho}=\Lambda_{\nu}{}^{a} \Lambda_{\rho}{}^{a}; \hspace{.2cm}v^{\mu}=
{\Sigma_0}^{\mu}\hspace{.3cm}
\label{spm2}
\end{equation}
where ${\Lambda_\mu}^{\alpha}$ is the inverse of $\Sigma_\alpha{}^\nu $, 
\begin{equation}
\Sigma_\alpha{}^\nu{\Lambda_\nu}^{\beta} =\delta^\beta_\alpha \hspace{.2cm};\hspace{.2cm}
\Sigma_\alpha{}^\nu{\Lambda_\mu}^{\alpha} =\delta^\nu_\mu \label{B}
\end{equation}

 So we find that the curved spacetime
obtained by GGT is the Newton Cartan spacetime \cite{BMM2}, \cite{BM3}.
\bigskip
\section{Galilean gauge theory as a non relativistic limit of Poincare gauge theory}
 We now delve into the new results of the paper. We will give a new derivation of Newtonian gravity as the non relativistic limit of GR without the need of introducing any extra conditions. Our method will be based on GGT which is able to provide a vielbien formulation of NC geometry. As far as we know, such a task was not successfully done earlier. %Cartan's original approach was to read off the connection coefficients from the equation of motion  of a particle in free fall, by comparing with the geodesic equation of the spacetime manifold. The method based on gauging the Poincare algebra depends on an Ansatz, which can hardly be motivated by physical reasoning. On the contrary, our approach does not assume anything extraneous.
 
  The method followed here essentially depends on two points. In the first place GR may be formulated as a tetrad formalism.
This  comes out to be identical with PGT. If PGT can be reduced to GGT in the nonrelativistic limit, then the same limit would allow the geometric dynamics of GR to pass to the
dynamics of gravity in NC geometry. There, however is a caveat. In the NC geometry the system always evolves along the flow of absolute time. So the last statement is true in the Galilean coordinates. GGT selects these coordinates by construction.

\subsection{Gauging the Poincare symmetry}
%In discussing Poincare symmetry we have to consider infinitesimal transformations of both fields $\phi(x)$ and the coordinates $x^{\mu}$. It is thus useful to consider two distinct variations. Form variations `$\delta_0$' that change the functional form at the same coordinates, $\delta_0 \phi(x)=\phi'(x)-\phi(x)$ and total variations which account for changes in both the functional form and the coordinates, $\delta \phi(x)=\phi'(x')-\phi(x)$. 

Consider a flat Minkowski space in any dimensions with metric $\eta_{\mu\nu}$. The Poincare group generators are composed of the angular momentum $L_{\mu\nu}=-x_{\mu}\partial_{\nu}+x_{\nu}\partial_{\mu}$, the spin $\Sigma_{\mu\nu}$ whose representation depends on the species of the field being acted upon and the translations $P_{\mu}=-\partial_{\mu}$. The first two are generally expressed in a combined form as $M_{\mu\nu}=L_{\mu\nu}+\Sigma_{\mu\nu}$, which is the total angular momentum.

The invariance of a Poincare symmetric theory is ensured by the fact that  the coordinates and fields tansform in a particular way. When the transformation parameters are localised by making these functions of space and time the invariance is lost. Clearly when the parameters are localised, local coordinates are to be introduced. The covariant derivatives are introduced in two steps.

The final covariant derivative is defined as \footnote{Greek indices from the beginning $\left(\alpha,\beta,....\right)$
denote the local basis while the global one is defined from the middle $\left(\mu,\nu,....\right)$},
\begin{align}
\nabla_\alpha\phi=\Sigma_\alpha{}^{\mu}\nabla_{\mu}\phi\label{covl}
\end{align}

Here, 
\begin{equation}
\nabla_{\mu}\phi=\partial_{\mu}\phi+\frac{1}{2}B^{\alpha\beta}_{\ \ \mu}\Sigma_{\alpha\beta}\phi\label{covdg1}
\end{equation}
We require $\nabla_\alpha\phi$ to transform
 covariantly under Poincare transformations, 
\begin{align}
\delta(\nabla_\alpha\phi)=-\frac{1}{2}\lambda^{\mu\nu}\Sigma_{\mu\nu}\nabla_{\alpha}\phi -\lambda_\alpha{}^\mu\nabla_{\mu}\phi\label{covltv}
\end{align}
For this, the form variations of the new fields must be
\begin{align}
\label{sigma}
\delta_0\Sigma_{\alpha}{}^{\mu} =-\xi^{\lambda}\partial_{\lambda}\Sigma_{\alpha}^{\mu}
+\partial_{\lambda}\xi^{\mu}\Sigma_{\alpha}{}^{\lambda} + \lambda_{\alpha}{}^{\beta}\Sigma_{\beta}{}^{\mu}
%\notag\\\delta_0\Sigma_0{}^0 &= -\xi^{\lambda}\partial_{\lambda}\Sigma_0{}^0
%+\partial_{\lambda}\xi^{\mu}\Sigma_{0}^{\lambda} + \lambda_0{}^{\alpha}\Sigma_{\alpha}{}^0 
\end{align}

and

\begin{align}
\delta_0 B^{\alpha\beta}{}_{\mu} &=- \partial_\mu\lambda^{\alpha\beta}-(\partial_\mu\xi^\lambda) B^{\alpha\beta}{}_{\lambda}-\xi^{\lambda}\partial_{\lambda}B^{\alpha\beta}{}_{\mu}
+\lambda^{\alpha\gamma} {B_{\gamma}}{}^{\beta}{}_{\mu}- \lambda^{\beta\gamma} 
{B_\gamma}{}^{\alpha}{}_{\mu}\label{delo} 
\end{align}

Having achieved the covariance of derivatives, we are now ready to define an invariant lagrangian {\textit{density}} $\tilde{{\cal{L}}}$ so that the action is invariant. One more correction is required to compensate the change in measure,
All these arguments eventually lead to a general form of the Poincare gauge theory invariant lagrangian as \cite{blago},
\begin{equation}
{\tilde{\mathcal{L}}}=b{\cal{L}}(\phi, \nabla_\alpha\phi)
\end{equation}
where $b$
is inverse of  $det\Sigma$.

It is possible to develop a geometric interpretation of this lagrangian that shows the identity of the transformattion laws (\ref{delo}) with those of the vielbeins and spin connection of Riemann Cartan space time. The curvature and torsion can be expressed in terms of the  vielbeins and spin connection coefficients. Imposing the restriction of vanishing torsion, GR can be formulated.
\subsection{Dictionary of fields connecting PGT with GGT}
The Poincare gauge theory contains 16 $\Sigma$ fields and 24 $B$ fields, The Galilean gauge theory, on the other hand, 
   contains 13 $\Sigma$ fields as $\Sigma_a^0 = 0$, The number of $B$ fields is same, We now proceed to prepare a complete dictionary to pass over from PGT to GGT. The implication of the identification will then be discussed,
  
From (\ref{sigma}) we can find the variation of $\Sigma_0{}^0 $
\begin{align}
\delta_0\Sigma_0{}^0 = -\xi^{\lambda}\partial_{\lambda}\Sigma_0{}^0
+\partial_{\lambda}\xi^{0}\Sigma_{0}^{\lambda} + \lambda_0{}^{\alpha}\Sigma_{\alpha}{}^0
\end{align}
Comparison with the corresponding GGT transformation (first equation of (\ref{crap1}))gives the conditions of matching
\begin{eqnarray}
\xi^{\mu}\to\zeta^{\mu}\nonumber\\
{\Sigma_a}^0 \to 0 
\end{eqnarray}
%\begin{table}[!ht]

Similarly the PGT transformation
\begin{equation}
\delta_0\Sigma_0{}^{k}= -\xi^{\nu}\partial_{\nu} \Sigma_0{}^k+\Sigma_0{}^{\nu}\partial_{\nu}\xi^{k}
+\lambda_{0}{}^a\Sigma_a{}^k
\end{equation}
maps to the corresponding  GGT transformation (second equation of (\ref{crap1})) provided the following  identification holds
\begin{eqnarray}
\lambda_0{}^{b}\to u^b
\end{eqnarray}

One can repeat the steps to establish complete matching of the PGT transformations
\begin{equation}
\delta_0\Sigma_a{}^k = -\xi^{\nu}\partial_{\nu}\Sigma_a{}^k+\Sigma_a{}^{\nu}
\partial_{\nu}\xi^k+ \lambda_a{}^0\Sigma_0{}^k+ \lambda_a{}^b\Sigma_b{}^k
\end{equation}
by making further correspondences,
%We see that reduction of the vielbein structures is completed with
\begin{eqnarray}
 \lambda_a{}^{b}\to -\omega_a{}^b\\
\lambda_a{}^{0} \to 0
\end{eqnarray}

This finishes the rules for reduction of the vielbeins of Riemann Cartan spacetime to those of the NC spacetime. The condition $\Sigma_a{}^0 = 0$ is an inbuilt condition of GGT. We can verify directly that this condition is consistent with PGT. Indeed using (\ref{sigma}) and the correspondences listed above, we can show that $\delta_0\Sigma_a{}^0 = 0$.

     Now the turn is of the spin connections. From PGT side, we can write using (\ref{delo})
     
\begin{align}
\delta_0 B_\mu{}^{a0} &=-\xi^\lambda\partial_\lambda B_\mu{}^{a0}-\partial_{\mu}\lambda^{a0}
-\partial_\mu\xi^\nu B_\nu{}^{a0} + \lambda^0{}_{b} B_\mu{}^{ab} + \lambda^a{}_{0} B_\mu{}^{00} + \lambda^a{}_{b} B_\mu{}^{b0}\notag\\
\delta_0 B_\mu{}^{ab}&=-\xi^\lambda\partial_\lambda B_\mu{}^{ab}-\partial_{\mu}\lambda^{ab}
-\partial_\mu\xi^\nu B_\nu{}^{ab} + \lambda^{b}{}_{0} B_\mu{}^{a0} +\lambda^{b}{}_{c} B_\mu{}^{ac}+\lambda^{a}{}_{0} B_\mu{}^{0b}+\lambda^{a}{}_{c} B_\mu{}^{cb} \label{ggtb}
\end{align}
%where use has been made of (\ref{crap}), 
For each coordinate index $\mu$ there are six independent spin connections, as $B_\mu {}^{\alpha\beta}$ is a second rank antisymmetric tensor. If we compare then with  ({\ref{crap}) complete match is found, if.
\begin{eqnarray}
B_\mu{}^{a0}\to B_\mu{}^{a}\\
 \lambda_b{}^{0}\to 0 \\ 
 \lambda^a{}_b\to \omega_b{}^a\\
 \lambda^{a0}\to-u^a 
\end{eqnarray}

%are required.
%\newpage

Further conditions are added when we consider $ B_\mu{}^{ab}$. Its variation in PGT is given by the second equation of (\ref{ggtb}). The reduction to  GGT requires
\begin{eqnarray}
B_\mu{}^{0b}\to 0\\
\lambda_b{}^{0}\to 0\\
\lambda^{ab}\to\omega^{ab}\\
\lambda^b{}_0\to 0
\end{eqnarray}
Incidentally, two of the above relations are also contained in the previous set.

For ready reference we compile the different correspondences  of the field components in table - 1. 
\begin{table}[!ht]
\begin{center}
\begin{tabular}{|c|c|c|c||c|c|}
\hline PGT & GGT & PGT & GGT & PGT & GGT\\ 
\hline $B_\mu{}^{a0}$ & $B_\mu{}^{a}$ & ${\Sigma_a}^0$ & $0$ & $B_\mu{}^{0a} $& $0$ \\
\hline
\end{tabular}
\label{chem1}
\caption{ Field Correspondences}
\end{center}
\end{table}
Clearly, these requirements are not surprising as these are the very conditions inbuilt in GGT. The conditions of the 
transformation parameters are listed in table - 2. The usefulness of these cannot be exaggerated . These correspondences enable us to convert any diffeomorphism invariant action in Riemann Cartan space with physics in the tangent space guided by STR to a  diffeomorphism invariant action in NC space time which satisfies galilean invariance in the flat limit. In particular this can be applied for gravity itself. This we will see in the next section.
\begin{table}[!ht]
\begin{center}
\begin{tabular}{|c|c|c|c|}
\hline PGT & GGT & PGT & GGT\\ 
\hline $\xi^{\mu}$ & $\zeta^{\mu}$ & $\lambda_0{}^{b}$ & $u^b$\\
\hline $\lambda_a{}^{b}$ & $-\omega_a{}^b$ &  $\lambda_a{}^{0}$ & $0$ \\ 
\hline $\lambda^0{}_b{}$ & $0$ & $\lambda^a{}_b$ & $-\omega_a{}^b$ \\
\hline $\lambda^{a0}$ & $-u^a$ &  $\lambda^{ab}$ & $\omega^{ab}$ \\
\hline $\lambda_b{}^{0}$ & $0$& $\lambda^{ 0a}$ &$0$ \\
%%\hline $\lambda^b{}_0$ & $0$ & $\lambda^{0a}$ & $0$ & $B_\mu{}^{0b}$ & $0$\\
\hline $\lambda^b{}_0$ & $0$ &  & \\
\hline
\end{tabular}
\caption{ Correspondences between the transformation parameters.}

\label{chem21}

\end{center}
\end{table}
\section{Newton's gravity from Galilean gauge theory}
In the previous section we have provided a dictionary to map the fields in PGT to those of GGT. The outcome is consistent with the physical ideas that we have already zeroed in while formulating GGT. For instance, we defined the covariant derivatives in a way that $\Sigma_a{}^0$ vanish. Similarly for the gauge field corresponding to galilean boost the structure proposed here differs markedly from its PGT counterpart ($B_\mu{}^{0a }=0$). Note that in the spatial subspace, the indices can be rasied or lowered, the parameter of transformations bear the same symmetry as in PGT. There is no such facility in the time space sector. This is due to the special role of time in galilean dynamics.

 The correspondence mentioned above opens up a possibility to reduce a theory in the Riemannian spacetime to Newton Cartan spacetime. Thus Einstein gravity can be reduced to its nonrelativistic form. At this point we must clearly understand that
 reduction to the Newton Cartan spacetime only is not sufficient. This Newton Cartan gravity has more degrees of freedom than Newton's gravity. To reduce to the latter the Galilean frame must be chosen where time is stratified by flat spatial surfaces
 piercing the time axis orthogonaly.

\subsection{The connection}
In the vielbein formalism the covariant derivatives are defined by the spin connections. 
The affine connection $\Gamma$ may be obtained by the vielbein postulate. Since we are considering torsionless theory $\Gamma$ must be symmetric. Thus,
\begin{equation}
\Gamma^{\rho}_{\nu\mu}=\frac{1}{2}\left(\partial_{\mu}\Lambda^{\alpha}{}_{\nu}\Sigma_{\alpha}{}^{\rho}+B^{\alpha}_{\mu\beta}\Lambda_{\nu}{}^{\beta}\Sigma_{\alpha}{}^{\rho}+\partial_{\nu}\Lambda^{\alpha}{}_{\mu}\Sigma_{\alpha}{}^{\rho}+B^{\alpha}_{\nu\beta}\Lambda_{\mu}{}^{\beta}\Sigma_{\alpha}{}^{\rho}\right)\label{g1}
\end{equation}
where $\Lambda$ is the inverse of $\Sigma$ i.e.
\begin{equation}
\label{X}
\Lambda^{\alpha}{}_{\nu}\Sigma_{\alpha}{}^{\rho} = \delta_\nu^\rho ;
\Lambda^{\alpha}{}_{\mu}\Sigma_{\beta}{}^{\mu} = \delta_\beta^\alpha 
\end{equation}
% We have earlier demonstrated that it is equivalent to the Dautcourt connection \cite{B}, the symmetric affine connection of Newton Cartan spacetime. Our aim is to calculate the elements of this  connection in the Galilean coordinates so that the Riemann tensor may be obtained in terms of the fields $\Sigma$ and $B$.

Putting $\rho = 0$ in the above equation, we get
\begin{equation}
2\Gamma^{0}_{\nu\mu}=\partial_{\mu}\Lambda^{\alpha}{}_{\nu}\Sigma_{\alpha}{}^{0}+B^{\alpha}_{\mu\beta}\Lambda_{\nu}{}^{\beta}\Sigma_{\alpha}{}^{0}+\partial_{\nu}\Lambda^{\alpha}{}_{\mu}\Sigma_{\alpha}{}^{0}+B^{\alpha}_{\nu\beta}\Lambda_{\mu}{}^{\beta}\Sigma_{\alpha}{}^{0}=0
\end{equation}
 Recalling  the fact that $\Sigma_a{}^0 = 0$ for all $a$ and using (\ref{X}), we get
%immediately shows
that for any $\mu$  and $\nu$,
\begin{equation}
\Gamma^0{}_{\nu\mu} = 0
\end{equation}

In our assumed frame, space is flat. So, $\Sigma_a{}^k =\delta_a{}^k$ should hold. Also ${\Lambda_k}^a = \delta_k^a$, which follows from (\ref{B}) and the special choice of ${\Sigma_k}^a$  .Using these results it is reassuring to note that the connection in the spatial subspace vanishes,
}
By a similar calculation we find 
\begin{eqnarray}
\Gamma^{k}_{l0} = \left(\partial_l\Lambda^a_0 + B_l{}^a\Lambda_0^0\right)\Sigma_a{}^k \label{GKL0}
\end{eqnarray}
%If we assume rectangular cartesian coordinates then only desired Newtonian equations will follow.
 In the galilean coordinates space is globally flat and time is absolute. So
$\Lambda_0^a$ cannot have a directional derivative in any spatial direction,
$$
\nabla_l\Lambda_0^a =0
$$
But,
\begin{eqnarray*}
\nabla_l\Lambda_0^a &=&  \left(\partial_l\Lambda^a_0 + B_l{}^{a0}\Lambda_0^0
+ B_l{}^{0a}\Lambda_a^0+ B_l{}^{ab}\Lambda_b^0
\right)\nonumber\\ &=& \left(\partial_l\Lambda^a_0 + B_l{}^a\Lambda_0^0\right)
\end{eqnarray*}
as $B_l{}^{ab} = 0$ and $ B_l{}^{0a} = 0$. Hence from ({\ref{GKL0}}), we get
\begin{eqnarray}
\Gamma^{k}_{l0} = \Big(\nabla_l\Lambda_0^a\Big)\Sigma_a{}^k = 0\label{GKL00}
\end{eqnarray}
It is worth metioning that ({\ref{GKL00}}) is a nontrivial result because none of the other methods could obtain this relation without assuming certain additional conditions.

% $The calculation has an interesting feature. After the usual cancellations we are left with 
%\begin{eqnarray}
%\Gamma^{k}_{l0} =  \partial_l\lambda^a{}_0\delta_a{}^k
%\end{eqnarray}
%The right hand side vanishes due to orthogonality of the foliations with evolution of time, as has been explained in theoutset.

 Finally, we compute the only nonzero component of the connection
\begin{eqnarray}
\Gamma^{k}_{00} =  \partial_0\Lambda_0{}^b\delta_b{}^k
+ B_0{}^a\delta_a{}^k\label{g}
\end{eqnarray}
To obtain the equations of Newton's gravity we write (\ref{g}) in terms of the gravitational potential $\Phi$ due to a static distribution of mass. Also the gravitational interaction propagates instantaneously. Note that we may then drop the time derivative of NC elements. In Newtonian gravity this static condition provides the rationale for dropping time derivatives.

 Introduce
\begin{eqnarray} 
B_0{}^a\delta_a{}^k = \partial^k\Phi\label{G1}
\end{eqnarray}
Then we get
\begin{eqnarray}
\Gamma^{k}_{00} =  \partial^k\Phi\label{gn}
\end{eqnarray}
This is an important result of our paper. Remember in Cartan's formulation, this result was read off from Newton's law of motion for a freely falling body. In other related formulations this result was obtained on using Trautman's condition along with certain global boundary conditions. Here their appearance as the outcome of our calculations is remarkable.

For our identification (\ref{gn}) it is crucial to show that
% can be shown from (\ref{g1})that 
 the left hand side transforms appropriately so that $\Phi$ is a 3-scalar.From (\ref{g1})and using (\ref{crap}),
 \begin{eqnarray}
\delta_0\left(B_0{}^a \delta_a{}^k\right) = - \omega^a_b B_{0}^{b}\delta_a{}^k + \partial_0 u^a\delta_a{}^k -\partial_0\zeta^\nu B_\nu{}^a \delta_a{}^k -\zeta^\nu\partial_\nu B_0{}^a \delta_a{}^k
\label{crapnn}
\end{eqnarray}
%Now we have to show that  transformation of the fields $\phi$ we consider apprpirate. 
Newtonian  gravity is static. Thus equation (\ref{crapnn}) becomes after 
dropping time derivatives,
\begin{eqnarray}
\delta_0\left(B_0{}^a \delta_a{}^k\right) = - \omega^a_b B_{0}^{b} \delta_a{}^k  -\zeta_\nu\partial^\nu B_0{}^a\delta_a{}^k
\label{crapnn1}
\end{eqnarray}

Now $\delta_a{}^k$ is a constant mixed tensor.
\begin{equation}
\delta_0\delta_a
^k =\left(\partial_m\zeta^k\delta_a
^m - \omega^{ab}\delta_b^k-\zeta^\nu\partial_\nu \delta_a{}^k\right)
\end{equation}
  Remembering the  constancy of delta we find that $\partial_m\zeta^k\delta_a
^m=\omega^{ab}\delta_b^k$. Now using it in (\ref{crapnn}) we get
\begin{eqnarray}
\delta_0\left(B_0{}^a \delta_a{}^k\right) =  \partial_m\zeta^k B_0{}^a \delta_a{}^m -\zeta_\nu\partial^\nu B_0{}^a\delta_a{}^k
\label{crapnn2}
\end{eqnarray}
This shows that $B_0{}^a \delta_a{}^k$ really transforms as a 3-space vector. But equation (\ref{gn}) then identifies $\Phi$ with a 3-scalar. Thus
we may conclude that $\Phi$ is the gravitational potential due to certain mass distribution. 

%Before passing to the next discussion, note that these connection coefficients are assumed as  axiom in the usual geometric formulation of Newtonian gravity. Other connection coefficients are assumed to be zero. Here their appearance as the outcome of our calculations is  remarkable. 
\subsection{Equations of Newtonian gravity}
In the Galileo Newton concept gravity is a force acting between two massive bodies. It is tacitly assumed that the interaction
propagates instantaneously. A certain mass distribution creates a conservative gravitational field. The strength of the gravitational field is obtained from the gravitatioal potential $\phi$, which satisfies Poisson's equation (\ref{poi.eqn}).
 This equation tells us how matter produces gravitational field.
If a test particle is subjected only to gravity it moves according to Newton's second law (\ref{ntp.eom}).
These equations  form Newton's theory of gravitation. Below, we will give their derivation as a geometric theory, by the reduction of Einstein's theory of gravitation according to the rules prescribed in the last section.

From the geometric point of view gravitation is not a force but curvature of spacetime, Test particles free of any (other) interaction move along geodesics defined by the equation (\ref{geo.eqn}).
Mass on the other hand influences the curvature of spacetime according to Einstein's equation
\begin{eqnarray}
R_{\mu\nu} - \frac{1}{2}{g_{\mu\nu}} R  = 8\pi G T_{\mu\nu}\label{eg}
\end{eqnarray}
where $g_{\mu\nu}$ is the metric, $ T_{\mu\nu}$ is the energy momentum tensor, $R_{\mu\nu}$ is Ricci tensor and $R$ is the Ricci scalar.

Einstein's theory is relativistic because it is formulated  
in Riemannian spacetime which has Lorentzian signature.It is well known \cite {blago} that this Riemannian manifol is generated by PGT. Using the correspondence  between PGT and GGT developed in the last section, we can write its non relativistic counterpart. Since
%and this holds irrespectuve of the fact whether the theory is relativistic or nonrelativistic  However some variables in our approach behaves differently in the respective cases.
 PGT offers a vielbein formulation of Einsteins gravity, we can then immediately write the dynamics in Newton Cartan space using GGT. Note that the Galileo Newton
 theory was formulated in flat space in cartesian coordinates. Hence, the geometric formulation will agree with it only in galilean coordinates. A vector, parallely transported along a closed curve suffers no change. The direction of the coordinate time coincides with the absolute time. Thus we  can conclude that 
\begin{enumerate}
\item
The vielbeins with all space indices may be replaced by Kronecker's
delta symbols.
\item
The spin connections with all space indices vanish.

\item
The geometric elements do not vary due to change of space coordinates.
\end{enumerate} 
 
\subsection{Equation of motion of a freely falling particle}
We first consider the motion of a test particle falling freely under gravity. According to the geometric theory of gravity, it follows the  geodesic equation (\ref{geo.eqn})in NC spacetime.
But we have proved that all components of $\Gamma^{\rho}_{\mu\nu}$ except $ \Gamma^{k}_{00}$ vanish. This only nonvanishing component is given by (\ref{gn}). Substitution immediately yields (\ref{ntp.eom})
which is the equation of motion of a freely falling test particle in the usual Galileo Newton formulation of the problem.

%Equation (\ref{mm}) gives the reaction of a test particle in a gravitational field. 

\subsubsection{Gravitational field due to a given mass distribution}
In a sense our task of deriving the equation (\ref{poi.eqn}) is complete with the deduction of all the components of the connection. From (\ref{gn}) one can switch to the usual definition of the Riemann tensor and show that the only non vanishing component of the Riemann tensor is $R_{00}$, where 
\begin{equation}
R_{00}=\nabla^2\Phi\label{r00}
\end{equation}
For nonrelativistic matter $T_{00} = T = \rho $ where $\rho$ is the mass density.
Subsituting all these in 
\begin{equation}
R_{\mu\nu}=8\pi G\left(T_{\mu\nu}-\frac{1}{2}Tg_{\mu\nu}\right)
\end{equation}
gives Poisson's equation (\ref{poi.eqn}).

A different route to arrive at (\ref{r00}), which depends entirely on the correspondence of PGT and GGT may, however, be interesting.

 The Riemann tensor is defined in PGT as \cite{blago}
\begin{equation}
R^{\mu}{}_{\nu\lambda\rho}=\Sigma_{\alpha}{}^{\mu}\Sigma_{\nu\beta}F^{\alpha\beta}{}_{\lambda\rho}\label{rie}
\end{equation}
where
\begin{equation}
F^{\alpha\beta}{}_{\lambda\rho}=\partial_{\lambda}B^{\alpha\beta}{}_{\rho}-\partial_{\rho}B^{\alpha\beta}{}_{\lambda}+B^{\alpha}{}_{\gamma\lambda}B^{\gamma\beta}{}_{\rho}-B^{\alpha}{}_{\gamma\rho}B^{\gamma\beta}{}_{\lambda}
\end{equation}
Using these, the Ricci tensor is obtained as,
\begin{equation}
R_{\nu\rho}=R^{\mu}{}_{\nu\mu\rho}=\Sigma_{\alpha}{}^{\mu}\Sigma_{\nu\beta}[\partial_{\mu}B^{\alpha\beta}{}_{\rho}-\partial_{\rho}B^{\alpha\beta}{}_{\mu}+B^{\alpha}{}_{\gamma\mu}B^{\gamma\beta}{}_{\rho}-B^{\alpha}{}_{\gamma\rho}B^{\gamma\beta}{}_{\mu}] \label{rt}
\end{equation}
and the Ricci scalar follows from it:
\begin{equation}
R=R^{\nu}{}_{\nu}=\Sigma_{\alpha}{}^{\mu}\Sigma_{\beta}{}^{\nu}[\partial_{\mu}B^{\alpha\beta}{}_{\nu}-\partial_{\nu}B^{\alpha\beta}{}_{\mu}+B^{\alpha}{}_{\mu\gamma}B^{\gamma\beta}{}_{\nu}-B^{\alpha}{}_{\gamma\nu}B^{\gamma\beta}{}_{\mu}]\label{rs}
\end{equation}.

So far our expressions were relativistic. Now we want to get the non relativistic limit. But this is easily accomplished by the method discussed in this paper.
We have already explained why it is necessary to stratify the Newton Cartan manifold and assume cartesian coordinates for obtaining Newtonian gravity as limit of Einstein's theory. The space is flat. We thus tabulate the necessary data

\begin{table}[!ht]
\begin{center}
\begin{tabular}{|c|c|}
\hline PGT & GGT \\ 
\hline $B_\mu{}^{a0}$ & $B_\mu{}^{a}$\\
\hline $\Sigma_a{}^{0}$ & $0$\\
%\hline $\Lambda^{a}{}_0$ & $0$\\
\hline
\end{tabular}
\label{chem22}
\end{center}
\end{table}
which is also contained as a subset of Table 1.

 Using the corresponding substitutions the different components of $ R_{\mu\nu}$ can be calculated.
Expcit calcularions give,$R_{kk}= 0$. Also

\begin{eqnarray}
R_{0k} = \delta_a^l\Sigma_{00}\left[\partial_l B_k{}^{a0} - \partial_k B_l{}^{a0}\right] \label{R0k}
\end{eqnarray}
The right hand side is actually zero in the frame chosen below. This can be proved as in the following. From the Vierbein postulate, we get,
\begin{align}
B_{\mu}{}^{\delta}{}_{\gamma}=\Gamma^{\rho}_{\mu\nu}\Sigma_{\gamma}{}^{\nu}\Lambda_{\rho}{}^{\delta}-\partial_{\mu}\Lambda^{\delta}{}_{\nu}\Sigma_{\gamma}{}^{\nu}
\end{align}
\begin{align}
B_k{}^a{}_0&=\partial_k\Lambda_{\nu}{}^a\Sigma_0{}^{\nu}
\end{align}
Note that $\Lambda_{k}{}^a$ is equal to $\delta_{k}{}^a$. Hence
 \begin{align}
B_k{}^a{}_0&=\partial_k\Lambda_{0}{}^a\Sigma_0{}^{0}= \partial_k\Lambda_{0}{}^a
\end{align}
Hence the right hand side of (\ref{R0k}) vanishes

We thus see that  the only non vanishing component of $ R_{\mu\nu}$ is $ R_{00}$ which is given by
\begin{eqnarray}
R_{00} = \partial_k\Big(\delta_a{}^kB_0{}^{a} \Big). 
\end{eqnarray}
Using (\ref{g1}) this reproduces  expression (\ref{r00}). As we have seen this result leads to Poisson's equation.
 This completes the derivation of Newtonian gravity on the Newton Cartan space.

\section{Conclusion}
In this paper we have established a detailed map between Poincare gauge theory (PGT) and Galilean gauge theory (GGT). Poincare gauge theory is known to provide a description of Riemann Cartan spacetime by the vielbeins and spin connections. Imposing symmetry of connection we then reach the spacetime of GR. On the other hand GGT gives a first order theory of NC geometry. The map between PGT and GGT thus enables us to connect a dynamical theory on Riemannian spacetime with that on 
NC spacetime.

As a specific application,  we have given a complete deduction of Newton's gravity as a geometric theory on Newton Cartan (NC) spacetime  from General Relativity (GR) on Riemannian spacetime, 
without assuming any  conditions or using any ansatze. Contrary to the earlier metric based approaches, ours is a vierbein approach.
In our analysis the connections are deduced from first principles, using the vierbein postulate. Incidentally the dynamical structure of the connection is such that only the gravitational potential appears and no additional ( Coriolis type) term shows up, as happens in other approaches \cite{Daut, TrautA, EHL}.

 One point should be noted in the above context. In the Galileo Newton concept,
 the motion takes place in a three dimensional Eucledian plane. According to the law of inertia this space can be covered by rectangular cartesian coordinares. Newton's laws of gravitation is formulated with respect to these coordinates.
 In this picture time is absolute and flows eternally. 	In the geometric picture
 this corresponds to a foliation  where the spatial leaves are flat and the coordinate time axis coincides with the direction of absolute time flow. GGT by construction adapts these coordinates.
 
  The algorithn presented here has wider applicability than deriving Newton's gravity from GR. Any complete dynamics on Riemann (or more generally, Riemann Cartan) spacetime can be reduced to its nonrelativistic form on NC space time
  using the maps derived here.
  This indicates the scope of future works.
 
 \section{Acknowledgement}
 One of the authors (PM) thanks the S. N. Bose National Centre for Basic Sciences for the short term visiting associate during which this work was completed.

\end{document}